\documentclass[12pt]{article}%
\usepackage{graphicx}
\usepackage{amsmath}
\usepackage{amsfonts}
\usepackage{amssymb}%
\providecommand{\U}[1]{\protect\rule{.1in}{.1in}}
\newtheorem{theorem}{Theorem}

\newtheorem{conjecture}[theorem]{Conjecture}

\newtheorem{definition}[theorem]{Definition}

\begin{document}

\title{\textbf{The critical temperature }$T_{cr}\left(  \text{Ising}\right)  $
\textbf{is DS-computable}}
\author{Senya Shlosman\\Krichever Center for Advance Studies, Moscow;\\Aix Marseille Univ, Universite de Toulon, \\CNRS, CPT, Marseille;\\Beijing Institute of Mathematical Sciences \\and Applications (BIMSA);\\Inst. of the Information Transmission Problems, \\RAS, Moscow\\shlosman@gmail.com}
\maketitle

\begin{abstract}
We show that the Dobrushin-Shlosman conditions $C_{V}$ for the uniqueness of
the Gibbs state provide the exact value for the critical temperature of the
$d$-dimensional Ising model.

\end{abstract}

\section{Introduction}

\subsection{DS uniqueness}

We begin with the reminder about the Dobrushin-Shlosman uniqueness conditions,
\cite{DS}. Consider a finite range lattice spin system $\sigma=\left\{
\sigma_{x},x\in\mathbb{Z}^{d}\right\}  ,$ defined by the (formal) Hamiltonian%
\[
H\left(  \sigma\right)  =-\sum_{A\subset\mathbb{Z}^{d}}U_{A}\left(
\sigma\right)  ,
\]
with single spins $\sigma_{x}$ taking values in a finite set $S$ and the
interaction $\mathcal{U}=\left\{  U_{A}\left(  \ast\right)  \right\}  ,$ where
the functions $U_{A}\left(  \sigma\right)  $ are translation-invariant
functions of $\sigma$ and depend only on the restriction $\sigma_{A}$ of the
configuration $\sigma$ to the set $A\subset\mathbb{Z}^{d};$ $U_{A}\equiv0$ for
$\mathrm{diam}\left(  A\right)  >R$. The nearest neighbor Ising model is the
best-known example.

One would like to know whether the interaction $\mathcal{U}$ defines the
corresponding Gibbs state $\mathbb{P},$
\[
\mathbb{P}\left(  \sigma\right)  \sim\exp\left\{  -H\left(  \sigma\right)
\right\}  ,
\]
in a unique way. The DS uniqueness conditions $C_{V},$ $V\subset\mathbb{Z}%
^{d},$ $\left\vert V\right\vert <\infty$ provide sufficient conditions for
that. The condition $C_{V}$ requires that the spins $\left\{  \sigma_{x},x\in
V\right\}  $ inside the box $V$ depend on the spins $\sigma_{y}$ outside $V$
weakly enough. It has the following form:

Let us fix the boundary condition $\sigma_{\partial V}=\left\{  \sigma
_{y},y\not \in V,\mathrm{dist}\left(  y,V\right)  \leq R\right\}  $ in the
region $\partial V$ around the finite box $V\subset\mathbb{Z}^{d}$ and
consider the conditional Gibbs distribution $q_{V}$ on spin configurations
$\sigma_{V}\in\Omega_{V}$ in $V:$
\[
q_{V}\left(  \sigma_{V}{\Huge |}\sigma_{\partial V}\right)  =\exp\left\{
-H_{V}\left(  \sigma_{V}{\Huge |}\sigma_{\partial V}\right)  \right\}
/Z\left(  V,\sigma_{\partial V}\right)  ,
\]
where%
\[
H_{V}\left(  \sigma_{V}{\Huge |}\sigma_{\partial V}\right)  =-\sum_{A\subset
V\cup\partial V}U_{A}\left(  \sigma_{V}{\Huge \cup}\sigma_{\partial V}\right)
,
\]
and
\[
Z\left(  V,\sigma_{\partial V}\right)  =\sum_{\sigma_{V}}\exp\left\{
-H_{V}\left(  \sigma_{V}{\Huge |}\sigma_{\partial V}\right)  \right\}
\]
is the partition function. Let $y\in\partial V$ be a site in $\partial V,$ and
we want to quantify the measure of dependence of the distribution
$q_{V}\left(  \ast{\Huge |}\sigma_{\partial V}\right)  $ on the value
$\sigma_{y}.$ So we write $\sigma_{\partial V}=\sigma_{\partial V\setminus
y}\cup\sigma_{y}$ and we consider a second boundary condition $\sigma
_{\partial V}^{\prime}=\sigma_{\partial V\setminus y}\cup\sigma_{y}^{\prime},$
which differs from $\sigma_{\partial V}$ at a single site $y.$ In that way we
get two probability distributions $q_{V}\left(  \ast{\Huge |}\sigma_{\partial
V}\right)  ,\ q_{V}\left(  \ast{\Huge |}\sigma_{\partial V}^{\prime}\right)
,$ and we want to measure how distinct they are. For that we endow the single
spin space $S$ with a metric $\rho$ and define the metric $\rho_{V}$ on
$\Omega_{V}$ by $\rho_{V}\left(  \sigma_{V},\bar{\sigma}_{V}\right)
=\sum_{x\in V}\rho\left(  \sigma_{x},\bar{\sigma}_{x}\right)  ,$ which allows
us to compute the \textbf{Kantorovich }(-Rubinstein-Ornstein-Waserstain-Monge)
\textbf{distance} $\mathfrak{d}_{K}$ between $q_{V}\left(  \ast{\Huge |}%
\sigma_{\partial V}\right)  ,\ q_{V}\left(  \ast{\Huge |}\sigma_{\partial
V}^{\prime}\right)  $. (We present the definition of $\mathfrak{d}_{K}$
below.) We define the value $k_{V,y}^{\mathcal{U}}$ -- the measure of the
dependence of the distribution $q_{V}\left(  \ast{\Huge |}\sigma_{\partial
V}\right)  $ on the value $\sigma_{y}$ -- by%
\[
k_{V,y}^{\mathcal{U}}=\max_{\sigma_{y}^{\prime}\neq\sigma_{y};\sigma_{\partial
V\setminus y}}\frac{\mathfrak{d}_{K}\left(  q_{V}\left(  \ast{\Huge |}%
\sigma_{\partial V\setminus y}\cup\sigma_{y}\right)  ,\ q_{V}\left(
\ast{\Huge |}\sigma_{\partial V\setminus y}\cup\sigma_{y}^{\prime}\right)
\right)  }{\rho\left(  \sigma_{y},\sigma_{y}^{\prime}\right)  }.
\]

\begin{definition}
$\left(  \cite{DS}\right)  $ Let $V\subset\mathbb{Z}^{d}$ be a finite box. The
interaction $\mathcal{U}$ satisfies the $C_{V}$ condition, if%
\begin{equation}
\sum_{y\in\partial V}k_{V,y}^{\mathcal{U}}<\left\vert V\right\vert .
\label{33}%
\end{equation}

\end{definition}

The main result of \cite{DS} is the following

\begin{theorem}
$\left(  \cite{DS}\right)  $ If for some finite $V\subset\mathbb{Z}^{d}$ the
interaction $\mathcal{U}$ satisfies the $C_{V}$ condition, then it has unique
Gibbs state. Moreover, this state has exponentially decaying correlations.
\end{theorem}

Denote by $\mathcal{A}_{V}$ the collection of all interactions $\mathcal{U}$
satisfying the condition $C_{V},$ and let $\mathcal{A}=\cup_{V\subset
\mathbb{Z}^{d}}\mathcal{A}_{V}.$ The (still open) conjecture of \cite{DS}
states the following:

\begin{conjecture}
For the interactions belonging to the boundary $\partial\mathcal{A}$ (some
king of) the phase transition takes place, and the surface $\partial
\mathcal{A}$ separates the interactions with and without phase transitions.
\end{conjecture}

The purpose of this work is to show that this conjecture holds for the
$d$-dimensional Ising model.\newline

\subsection{Main result}

The Ising model at the inverse temperature $\beta$ is given by the Hamiltonian%
\[
H^{\beta}\left(  \sigma\right)  =-\beta\sum_{\substack{x,y\in\mathbb{Z}%
^{d}\\\left\vert x-y\right\vert =1}}\sigma_{x}\sigma_{y},
\]
$\sigma_{x},\sigma_{y}=\pm1.$ (We include the inverse temperature $\beta$ into
the Hamiltonian.) For every $V$ define $\beta_{V}$ to be the maximal value
such that for $\beta<\beta_{V}$ the condition $C_{V}$ holds for $H^{\beta}.$
Our main result claims that
\begin{equation}
\sup_{V}\beta_{V}=\beta_{cr}\left(  d\right)  , \label{21}%
\end{equation}
where $\beta_{cr}\left(  d\right)  $ is the inverse critical temperature of
the $d$-dimensional Ising model. The critical temperature $T_{cr}\left(
d\right)  \equiv\beta_{cr}^{-1}\left(  d\right)  $ is the lowest temperature
above which the pair correlation function $\left\langle \sigma_{x}\sigma
_{y}\right\rangle ^{\beta}$ decays exponentially:%
\[
\left\langle \sigma_{x}\sigma_{y}\right\rangle ^{\beta}\leq\exp\left\{
-c\left(  \beta\right)  \left\vert x-y\right\vert \right\}  ,
\]
where $\sigma_{x},\sigma_{y}=\pm1$ are the Ising spins, $x,y\in\mathbb{Z}%
^{d},$ $\left\langle \ast\right\rangle ^{\beta}$ is the (infinite volume)
$\left(  +\right)  $-state of the $d$-dimensional Ising model at the
temperature $T=\beta^{-1},$ and $c\left(  \beta\right)  >0$ is some constant.

Historically, the first known condition on the temperature $T,$ implying that
$T$ is above the criticality is the Dobrushin uniqueness condition, \cite{D}.
In fact, it coincides with the simplest condition among $C_{V}$-s, which
corresponds to the case $\left\vert V\right\vert =1.$ To check the Dobrushin
uniqueness condition for the model with nearest-neighbor interaction (i.e.
$R=1$) one needs to know only the family of conditional distributions of a
single spin $\sigma_{x}$ under condition that all the $2d$ neighboring spin
variables $\sigma_{y},$ $\left\vert x-y\right\vert =1,$ are fixed. In the case
of the Ising model these conditional distributions are given by%
\begin{align*}
&  q_{x}^{\beta}\left(  \sigma_{x}{\Huge |}\sigma_{y},y\in\partial\left\{
x\right\}  \right) \\
&  =\exp\left\{  \beta\sum_{y:\left\vert x-y\right\vert =1}\sigma_{x}%
\sigma_{y}\right\}  \left[  \sum_{\sigma_{x}=\pm1}\exp\left\{  \beta
\sum_{y:\left\vert x-y\right\vert =1}\sigma_{x}\sigma_{y}\right\}  \right]
^{-1},
\end{align*}
they are indexed by the configurations $\sigma_{\partial\left\{  x\right\}  }$
of $2d$ spins $\left\{  \sigma_{y},y\in\partial\left\{  x\right\}  \right\}
$. One has to determine the dependence $k_{x,y}^{\beta}$ of this distribution
on the spin at $y\in\partial\left\{  x\right\}  ,$ maximized over all possible
values of the conditioning $\sigma_{\partial\left\{  x\right\}  \setminus y}.$
Dobrushin uniqueness condition then reads:%
\begin{equation}
\sum_{y:\left\vert x-y\right\vert =1}k_{x,y}^{\beta}<1. \label{01}%
\end{equation}
Since by symmetry all the $2d$ values $k_{x,y}^{\beta}$ are the same, one can
likewise formulate it as $k_{x,y}^{\beta}<\frac{1}{2d}.$ (For example,
$k_{x,y}^{\beta=0}=0.$) Let us define the inverse temperature $\beta
_{D}\left(  d\right)  $ to be the $\sup\left\{  \beta:k_{x,y}^{\beta}<\frac
{1}{2d}\right\}  .$ Then, as Dobrushin has shown in \cite{D}, $\beta
_{D}\left(  d\right)  <\beta_{cr}\left(  d\right)  ,$ and one has uniqueness
for all $\beta<\beta_{D}\left(  d\right)  .$

Of course, the gap between the inverse temperatures $\beta_{D}\left(
d\right)  $ and $\beta_{cr}\left(  d\right)  $ is substantial. Our
\textbf{Main Result} is that this gap is filled completely by the help of
$\beta_{V}$-s:

\begin{theorem}
\label{M} The uniqueness conditions $C_{V}$ exhaust all the supercritical
temperatures of the $d$-dimensional Ising model:%
\begin{equation}
\sup\left\{  \beta_{V}\left(  d\right)  ,V\subset\mathbb{Z}^{d}\right\}
=\beta_{cr}\left(  d\right)  . \label{22}%
\end{equation}

\end{theorem}

Our proof is based on the following Ding-Song-Sun correlation inequality,
\cite{DSS}. Consider the Ising model in the finite box $V$ with empty boundary
conditions, at inverse temperature $\beta$ and under magnetic field
$\mathbf{h}=\left\{  h_{x},x\in V\right\}  .$ Let $u,v\in V$ be two sites, and
consider the covariance, $\left\langle \sigma_{u};\sigma_{v}\right\rangle
^{\mathbf{h}}=\left\langle \sigma_{u}\sigma_{v}\right\rangle ^{\mathbf{h}%
}-\left\langle \sigma_{u}\right\rangle ^{\mathbf{h}}\left\langle \sigma
_{v}\right\rangle ^{\mathbf{h}}.$ It turns out that the covariance is maximal
when all $h$-s are zero:
\begin{equation}
\left\langle \sigma_{u};\sigma_{v}\right\rangle ^{\mathbf{h}}\leq\left\langle
\sigma_{u};\sigma_{v}\right\rangle ^{\mathbf{h}=\mathbf{0}}\equiv\left\langle
\sigma_{u}\sigma_{v}\right\rangle ^{\mathbf{h}=\mathbf{0}}. \label{03}%
\end{equation}
For the case of the fields $\mathbf{h}$ being positive, the inequality
$\left(  \ref{03}\right)  $ is a special case of the GHS inequality. But the
general case was open for many years, though several people (including the
author of the present paper) conjectured it to be true for all values of
$\mathbf{h}$.\newline

\subsection{The critical temperature $T_{cr}\left(  d\right)  $ is
constructive}

The relation $\left(  \ref{22}\right)  $ implies that the critical temperature
$T_{cr}\left(  d\right)  $ can be in principle computed to arbitrary precision
by evaluating the temperatures $T_{V}$-s, above which the corresponding
condition $C_{V}$ holds. Since checking $C_{V}$ requires a finite amount of
computations, involving only elementary functions, we say that the critical
temperature $T_{cr}\left(  d\right)  $ of the Ising model is constructive. It
is not hard to implement the relevant computations and check some conditions
$C_{V}$. One example is considered in \cite{DKS}; the box $V$ used there was a
$3\times4$ rectangle.

The constructivity of the Ising $T_{cr}\left(  d\right)  $ was discovered
earlier by H. Duminil-Copin and V. Tassion, \cite{DCT}. They have constructed
(computable) functions $\varphi_{\beta}\left(  V\right)  $ (for the Ising
model) and $\varphi_{p}\left(  V\right)  $ (for the Bernoulli percolation with
parameter $p$) with the properties:

\textbf{1. }If $\varphi_{\beta}\left(  V\right)  <1$ for some finite box $V$
then $\beta<\beta_{cr}$. If $\varphi_{p}\left(  V\right)  <1$ for some finite
$V$ then there is no infinite open cluster in $p$-Bernoulli percolation.

\textbf{2. }If $\varphi_{\beta}\left(  V\right)  \geq1$ for all finite boxes
$V$ then $\beta\geq\beta_{cr}$. If $\varphi_{p}\left(  V\right)  \geq1$ for
all finite $V$ then for each $\tilde{p}>p$ with probability one there is an
infinite open cluster in $\tilde{p}$-Bernoulli percolation.\newline

\section{Constructive uniqueness}

In this section we remind the reader the definition of the dependence
coefficients $k_{V,y}^{\beta}$ entering $\left(  \ref{33}\right)  .$

\subsection{KROV distance}

Let $\left(  X,\rho\right)  $ be a metric space, and $\mu,\nu$ be two
probability measures on $X.$ A probability measure $P$ on $X\times X$ is
called a coupling of $\mu$ and $\nu$ iff for any $A\subset X$
\[
P\left(  A\times X\right)  =\mu\left(  A\right)  ,\ P\left(  X\times A\right)
=\nu\left(  A\right)  .
\]
The set of all such couplings is denoted by $\Pi\left(  \mu,\nu\right)  .$ The
Kantorovich-Rubinstein-Ornstein-Wasserstein distance $\mathfrak{d}_{K}\left(
\mu,\nu\right)  $ is defined by%
\[
\mathfrak{d}_{K}\left(  \mu,\nu\right)  =\inf_{P\in\Pi\left(  \mu,\nu\right)
}\int_{X\times X}\rho\left(  x,y\right)  P\left(  dx,dy\right)  .
\]
For a special case when $\rho\left(  x,y\right)  =1$ for any $x\neq y$, the
distance $\mathfrak{d}_{K}\left(  \mu,\nu\right)  $ is the same as the
variational distance $var\left(  \mu,\nu\right)  .$

\subsection{Dependence coefficients}

Let $V\subset\mathbb{Z}^{d}$ be a finite box, $\Omega_{V}=\left\{  \sigma
_{V}=\left\{  \sigma_{x}=\pm1:x\in V\right\}  \right\}  $ be the set of all
Ising spin configurations in $V,$ and $\partial V=\left\{  y\notin
V:\mathrm{dist}\left(  y,V\right)  =1\right\}  .$ The Ising model energy of a
configuration $\sigma_{V}\in\Omega_{V}$ with boundary condition $\sigma
_{\partial V}\in\Omega_{\partial V}$ is given by
\[
H_{V}\left(  \sigma_{V}{\LARGE |}\sigma_{\partial V}\right)  =-\sum_{x\sim
y\in V}\sigma_{x}\sigma_{y}-\sum_{\substack{x\sim y\\x\in V,y\in\partial
V}}\sigma_{x}\sigma_{y},
\]
where $x\sim y$ denotes the nearest neighbors. The conditional Gibbs
distribution of the Ising model in the box $V$ at inverse temperature $\beta$
given the boundary condition $\sigma_{\partial V}\in\Omega_{\partial V}$ is
the following probability distribution on $\Omega_{V}:$%
\[
q_{V}^{\beta}\left(  \sigma_{V}{\LARGE |}\sigma_{\partial V}\right)
=\exp\left\{  -\beta H_{V}\left(  \sigma_{V}{\LARGE |}\sigma_{\partial
V}\right)  \right\}  {\Huge /}Z\left(  V,\beta,\sigma_{\partial V}\right)  ,
\]
where the partition function $Z\left(  V,\beta,\sigma_{\partial V}\right)  $
is given by%
\[
Z\left(  V,\beta,\sigma_{\partial V}\right)  =\sum_{\sigma_{V}\in\Omega_{V}%
}\exp\left\{  -\beta H_{V}\left(  \sigma_{V}{\LARGE |}\sigma_{\partial
V}\right)  \right\}  .
\]
We introduce the metric $\rho$ on $\Omega_{V}$ via%
\[
\rho\left(  \sigma_{V}^{\prime},\sigma_{V}^{\prime\prime}\right)  =\sum_{x\in
V}\left\vert \sigma_{x}^{\prime}-\sigma_{x}^{\prime\prime}\right\vert .
\]

Let $y\in\partial V;$ for a boundary condition $\sigma_{\partial V}\in
\Omega_{\partial V}$ we define its flip $\sigma_{\partial V}^{y}$ at $y$ to be
a configuration obtained from $\sigma_{\partial V}$ by changing its value at a
single point $y,$ from $\sigma_{y}$ to $-\sigma_{y}.$ Now we can compute the
distance $\mathfrak{d}_{K}\left(  q_{V}^{\beta}\left(  \ast{\LARGE |}%
\sigma_{\partial V}\right)  ,q_{V}^{\beta}\left(  \ast{\LARGE |}%
\sigma_{\partial V}^{y}\right)  \right)  ,$ and finally define the dependence
coefficients $k_{V,y}^{\beta}$ via%
\[
k_{V,y}^{\beta}=\frac{1}{2}\max_{\sigma_{\partial V}\in\Omega_{\partial V}%
}\mathfrak{d}_{K}\left(  q_{V}^{\beta}\left(  \ast{\LARGE |}\sigma_{\partial
V}\right)  ,q_{V}^{\beta}\left(  \ast{\LARGE |}\sigma_{\partial V}^{y}\right)
\right)  .
\]
(The factor $\frac{1}{2}$ in the definition comes from the fact that
$\left\vert \sigma_{x}-\left(  \sigma_{x}^{x}\right)  \right\vert =2.$)

The main result of \cite{DS} applied to the Ising model is the following.

\begin{theorem}
\label{main} Suppose that for some finite box $V$ and inverse temperature
$\beta$ the condition $C_{V}$ holds:%
\[
\sum_{y\in\partial V}k_{V,y}^{\beta}<\left\vert V\right\vert .
\]
Then the Ising model on $\mathbb{Z}^{d}$ at inverse temperature $\beta$ has
unique Gibbs state, with exponentially decaying correlations.
\end{theorem}

\subsection{Proof of the Theorem \ref{M}.}

Let $V$ be a finite box, $y\in\partial V,$ and $\sigma_{\partial V}\in
\Omega_{\partial V}$ -- some boundary condition. Consider the optimal coupling
$P\in\Pi\left(  q_{V}^{\beta}\left(  \ast{\LARGE |}\sigma_{\partial V}\right)
,q_{V}^{\beta}\left(  \ast{\LARGE |}\sigma_{\partial V}^{y}\right)  \right)  $
between the measures $q_{V}^{\beta}\left(  \ast{\LARGE |}\sigma_{\partial
V}\right)  $ and $q_{V}^{\beta}\left(  \ast{\LARGE |}\sigma_{\partial V}%
^{y}\right)  $ on $\Omega_{V}.$ It means that
\[
\int_{\Omega_{V}\times\Omega_{V}}\sum_{x\in V}\left\vert \sigma_{x}^{\prime
}-\sigma_{x}^{\prime\prime}\right\vert P\left(  \sigma_{V}^{\prime},\sigma
_{V}^{\prime\prime}\right)  =\mathfrak{d}_{K}\left(  q_{V}^{\beta}\left(
\ast{\LARGE |}\sigma_{\partial V}\right)  ,q_{V}^{\beta}\left(  \ast
{\LARGE |}\sigma_{\partial V}^{y}\right)  \right)  .
\]
Let us assume that the boundary condition configuration $\sigma_{\partial V}$
is $+1$ at $y,$ so $\sigma_{\partial V}^{y}$ is $-1$ at $y.$ Then, by
definition, $\sigma_{\partial V}$ is \textit{higher} than $\sigma_{\partial
V}^{y}.$ Therefore there exists a \textit{monotone coupling} $P^{FKG}\in
\Pi\left(  q_{V}^{\beta}\left(  \ast{\LARGE |}\sigma_{\partial V}\right)
,q_{V}^{\beta}\left(  \ast{\LARGE |}\sigma_{\partial V}^{y}\right)  \right)  $
such that $P^{FKG}\left(  \sigma_{V}^{\prime},\sigma_{V}^{\prime\prime
}\right)  >0$ implies that $\left(  \sigma_{V}^{\prime}\right)  _{x}%
\geq\left(  \sigma_{V}^{\prime\prime}\right)  _{x}$ at every site $x\in V.$
(See e.g. \cite{H} for the details concerning the Fortuin-Kastelein-Ginibre
coupling $P^{FKG}$.) Hence%
\begin{align*}
\int_{\Omega_{V}\times\Omega_{V}}\sum_{x\in V}\left\vert \sigma_{x}^{\prime
}-\sigma_{x}^{\prime\prime}\right\vert P\left(  \sigma_{V}^{\prime},\sigma
_{V}^{\prime\prime}\right)   &  \leq\int_{\Omega_{V}\times\Omega_{V}}%
\sum_{x\in V}\left\vert \sigma_{x}^{\prime}-\sigma_{x}^{\prime\prime
}\right\vert P^{FKG}\left(  \sigma_{V}^{\prime},\sigma_{V}^{\prime\prime
}\right) \\
&  =\int_{\Omega_{V}\times\Omega_{V}}\sum_{x\in V}\left(  \sigma_{x}^{\prime
}-\sigma_{x}^{\prime\prime}\right)  P^{FKG}\left(  \sigma_{V}^{\prime}%
,\sigma_{V}^{\prime\prime}\right) \\
&  =\sum_{x\in V}\left(  \left\langle \sigma_{x}\right\rangle _{V,\sigma
_{\partial V}}^{\beta}-\left\langle \sigma_{x}\right\rangle _{V,\sigma
_{\partial V}^{y}}^{\beta}\right)  ,
\end{align*}
where $\left\langle \ast\right\rangle _{V,\sigma_{\partial V}}^{\beta}$ is the
expectation with respect to the measure $q_{V}^{\beta}\left(  \ast
{\LARGE |}\sigma_{\partial V}\right)  ,$ etc.

Let us now consider the box $V\cup y$ with boundary condition $\sigma
_{\partial V\setminus y}$ and define the magnetic field $h=h\left(
V,\sigma_{\partial V\setminus y}\right)  $ at $y$ (and zero at all other
points) to be the value which makes the expectation $\left\langle \sigma
_{y}\right\rangle _{V\cup y,\sigma_{\partial V\setminus y}}^{\beta,h}$ to
vanish. Then the difference $\left\langle \sigma_{x}\right\rangle
_{V,\sigma_{\partial V}}^{\beta}-\left\langle \sigma_{x}\right\rangle
_{V,\sigma_{\partial V}^{y}}^{\beta}$ is nothing else but the covariance
$\left\langle \sigma_{x};\sigma_{y}\right\rangle _{V\cup y,\sigma_{\partial
V\setminus y}}^{\beta,h}$ (up to a factor $2$). If $\beta<\beta_{cr},$ then,
due to the inequality $\left(  \ref{03}\right)  ,$ the covariance
$\left\langle \sigma_{x};\sigma_{y}\right\rangle _{V\cup y,\sigma_{\partial
V\setminus y}}^{\beta,h}$ is upper bounded by $\exp\left\{  -c\left(
\beta\right)  \left\vert x-y\right\vert \right\}  $ for some $c\left(
\beta\right)  >0.$ $\blacksquare$

\textbf{Acknowledgement. }\textit{The author would like to thank Professor
Jian Ding for helpful discussion. The work was supported by the RSF under
project 23-11-00150.}

\end{document}